\documentclass{article}
\usepackage{spconf,amsmath,graphicx}

\newlength\savedwidth
\newcommand{\wcline}[1]{\noalign{\global\savedwidth\arrayrulewidth\global\arrayrulewidth 1.0pt} \cline{#1}
\noalign{\global\arrayrulewidth\savedwidth}}
\usepackage{multirow}
\usepackage{amsmath}
\usepackage{amsfonts}
\usepackage{enumerate}
\usepackage{colortbl}
\definecolor{mycolor1}{cmyk}{0.99,0.0,0.81,0.0}
\definecolor{mycolor2}{cmyk}{0.08,0.0,0.09,0.19}
\usepackage{wrapfig}
\usepackage{caption}
\usepackage{cite}
\usepackage{url}
\usepackage{here}
\usepackage{comment}
\usepackage{arydshln}
\usepackage{setspace}

\makeatletter
\def\bstctlcite{\@ifnextchar[{\@bstctlcite}{\@bstctlcite[@auxout]}}
\def\@bstctlcite[#1]#2{\@bsphack
\@for\@citeb:=#2\do{%
\edef\@citeb{\expandafter\@firstofone\@citeb}%
\if@filesw\immediate\write\csname #1\endcsname{\string\citation{\@citeb}}\fi}%
\@esphack}
\makeatother

\title{Sound event detection based on curriculum learning \\considering learning difficulty of events}
 \name{Noriyuki Tonami$^{1}$,
       Keisuke Imoto$^{2}$,
       Yuki Okamoto$^{1}$,
       Takahiro Fukumori$^{1}$, 
       Yoichi Yamashita$^{1}$
       }
 \address{$^1$ Ritsumeikan University, Japan, \ $^2$ Doshisha University, Japan
}

\begin{document}

\maketitle
\begin{abstract}
In conventional sound event detection (SED) models, two types of events, namely, those that are present and those that do not occur in an acoustic scene, are regarded as the same type of events.
The conventional SED methods cannot effectively exploit the difference between the two types of events.
All time frames of sound events that do not occur in an acoustic scene are easily regarded as inactive in the scene, that is, the events are easy-to-train.
The time frames of the events that are present in a scene must be classified as active in addition to inactive in the acoustic scene, that is, the events are difficult-to-train.
To take advantage of the training difficulty, we apply curriculum learning into SED, where models are trained from easy- to difficult-to-train events.
To utilize the curriculum learning, we propose a new objective function for SED, wherein the events are trained from easy- to difficult-to-train events.
Experimental results show that the F-score of the proposed method is improved by 10.09 percentage points compared with that of the conventional binary cross entropy-based SED.

\end{abstract}
\vspace{-3pt}
\section{Introduction}
\label{sec:intro}
\vspace{-3pt}

The analysis of various environmental sounds in everyday life has be come an increasingly important area in signal processing \cite{Imoto_AST2018_01}.
The automatic analysis of environmental sounds will give rise to various applications, such as anomalous sound detection systems \cite{abnormal}, automatic life-logging systems \cite{Stork_ROMAN2012_01}, monitoring systems \cite{Ntalampiras_ICASSP2009_01}, and bird-call detection systems  \cite{Okamoto_crow}.

Sound event detection (SED) is the task of recognizing sound event labels and their timestamp from a recording.
In SED, the models need to recognize overlapped multiple sound events in a time frame.
Recently, neural-network-based SED models have seen increasingly rapid advances, such as the convolutional neural network (CNN) \cite{Hershey_ICASSP2017_01}, recurrent neural network (RNN) \cite{Hayashi_TASLP2017_01}, and convolutional recurrent neural network (CRNN) \cite{SED_CRNN}.
CNN is the structure that automatically extracts features and is robust to time and frequency shifts.
RNN is good at modeling the time structure in an audio stream.
Moreover, some works considering the relationship between sound events and scenes have been proposed.
As an example of the relationship, ``mouse clicking'' occurs indoors such as ``office,'' whereas, ``car'' tends to occurs outdoor such as ``city center.''
On the basis of this idea, SED using the information on the acoustic scene  \cite{Mesaros_EUSIPCO2011_01,Heittola_JASM2013_01,Imoto_IEICE2016_01} and the model combining SED and acoustic scene classification (ASC) \cite{helen_interspeech2019,Tonami_WASPAA2019_01,imoto_tonami_icassp2020,Komatsu_icassp2020,dcasenet_arxiv} have been proposed. 
Heittola $\textit{et al}$. \cite{Heittola_JASM2013_01} have proposed the SED model using the results of the ASC, where the ASC model is trained in the first stage and then the SED model is trained in the second stage with the ASC results.
Tonami $\textit{et al}$. \cite{Tonami_WASPAA2019_01} have proposed the multitask-learning-based models combining SED and ASC.

\begin{figure}[t!]
  \centering
  \includegraphics[width=1.00\columnwidth]{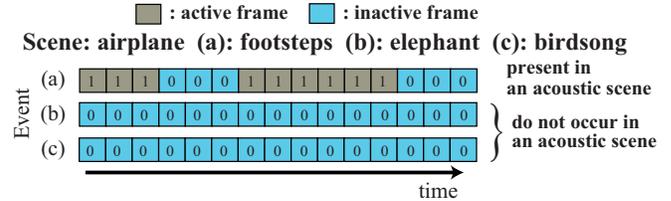}
  \vspace{-20pt}
  \captionsetup{labelformat=empty}
  \caption{{\bf Fig. 1}. Example of difference in training difficulty between sound events}
  \label{fig:difference_difficulty}
  \vspace{-10pt}
\end{figure}

\begin{figure*}[t!]
  \centering
  \includegraphics[width=2.00\columnwidth]{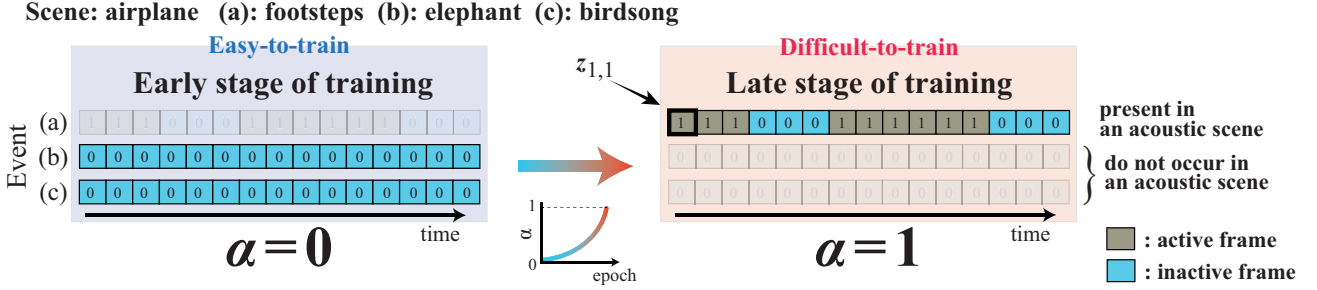}
  \captionsetup{labelformat=empty}
  \caption{{\bf Fig. 2}. Examples of early and late stage of training based on curriculum learning}
  \label{fig:state_learning}
  \vspace{-2pt}
\end{figure*}

In the conventional SED methods, two types of events, namely, those that are present and those that do not occur in an acoustic scene, are treated as the same type of the events.
The conventional SED methods cannot effectively utilize the difference between the two types of events.
The all time frames of events that do not occur in a scene only need to be treated as inactive in the acoustic scene, as shown in Fig. \ref{fig:difference_difficulty} (``elephant'' and ''birdsong'' in ``airplane''), i.e., the training of the easy-to-train events is considered as the task of recognizing one class.
On the other hand, the time frames of events that are present in an acoustic scene must be classified as active or inactive in the acoustic scene, as shown in Fig. \ref{fig:difference_difficulty} (``footsteps'' in ``airplane''), i.e., the training of the difficult-to-train events is regarded as the task of binary classification.

To utilize the difference in the difficulty of training between the sound events, we employ curriculum learning \cite{curriculum}.
Curriculum learning is a method of learning data effectively considering the difficulty of training, in which a model learns  progressively from easy- to difficult-to-train data.
Recently, some works using the curriculum learning have been carried out \cite{Speech_noise_curriculum,SER_curriculum,Speech_trains_curriculum}. 
Lotfian and Busso \cite{SER_curriculum} have proposed the speech emotion recognition method based on the curriculum learning, where the ambiguity of emotion is considered.
In this paper, we propose a SED method using the curriculum learning, in which strong labels are given for the training.
In the proposed method, the SED models are trained from the easy- to difficult-to-train events on the basis of the curriculum learning.
More specifically, we present a new objective function of SED considering the difficulty of the training of events based on the curriculum learning. 
%
%
\vspace{-3pt}
\section{Conventional method}
\label{sec:conv}
\vspace{-3pt}
SED involves sound event labels and their onset/offset from an audio.
Recently, many neural-network-based methods have been studied.
In most of the neural-network-based methods, the acoustic features in the time-frequency domain are used for the input to the SED models.
To optimize the neural-network-based SED models, the binary cross-entropy loss is used as follows: 

\vspace{-10pt}   
\begin{align}
\hspace*{0pt} {\mathcal L}_{\rm BCE} 
 = - \! \sum^{N}_{n=1} \sum^{T}_{t=1} \! {\Big \{} z_{n,t} \log {\sigma}(y_{n,t}) \! 
\nonumber\\[-1pt]
+ ~\! (1 \! - \! z_{n,t}) &\log {\big (} 1 \! - \! {\sigma}(y_{n,t}) {\big )} {\Big \}},
\label{eq:event_loss}
\vspace{-20pt}
\end{align}

\noindent where {\it N} and {\it T} indicate the numbers of sound event categories and time frames, respectively.
$z_{n,t} \in \{ 0,1 \}$ is a target label of an event {\it n} at time {\it t}.
If the event is active, $z_{n,t}$ is 1; otherwise, $z_{n,t}$ is 0.
$y_{n,t}$ represents the output of the network of an event {\it n} at time {\it t}. 
$\sigma(\cdot)$ denotes the sigmoid function.

\vspace{-5pt}
\section{Proposed method}
\vspace{-3pt}
\vspace{0pt}
\subsection{Training difficulty of events considering scenes}
\vspace{-3pt}
%
In the conventional SED methods, two types of events, namely, those that exist and those that do  not occur in an acoustic scene, are treated as the same type of the events.
The conventional SED methods cannot effectively employ the difference between the two types of sound events.
The all time frames of events that do not occur in an acoustic scene only need to be regarded as inactive in the acoustic scene, as seen in Fig. \ref{fig:difference_difficulty} (``elephant'' and ``birdsong'' in ``airplane''). 
The training of the sound events is treated as the task that recognizes one class (inactive), that is, the events are easy-to-train.
On the other hand, the time frames of sound events that exist in an acoustic scene need to be classified as active in addition to inactive in the acoustic scene, as shown in Fig. \ref{fig:difference_difficulty}.
The training of the sound events is considered as the task that classifies two classes (active or inactive), that is, the events are difficult-to-train.
In short, the sound events that exist in an acoustic scene are hardly trained compared with the events that do not occur in the acoustic scene as shown in Fig. \ref{fig:state_learning}.
\begin{figure*}[t!]
  \centering
  \includegraphics[width=2.0\columnwidth]{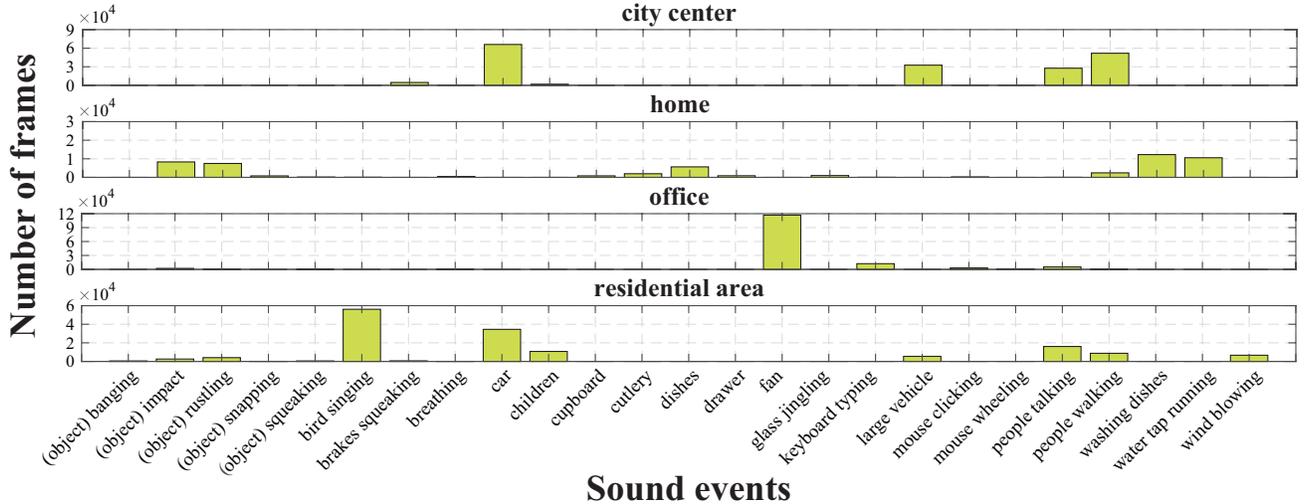}
  \vspace{-10pt}
  \captionsetup{labelformat=empty,labelsep=none}
  \caption{{\bf Fig.~3}. Number of frames of sound events on development set used for our experiments}
  \label{fig:frames_events}
  \vspace{-5pt}
\end{figure*}
%
\subsection{Curriculum-learning-based objective function}
\label{sec:prop}
\vspace{-3pt}
As mentioned in Sect. 3.1., there are differences in training difficulty between the sound events when the acoustic scenes are considered. 
In the proposed method, we employ the curriculum learning to take advantage of the difference in the difficulty of training between the sound events when the acoustic scenes are considered.
To incorporate the concept of the curriculum learning into the BCE, the following loss function is used instead of Eq. \ref{eq:event_loss}:

\vspace{-10pt}
\begin{align}
\vspace{-25pt}
\hspace*{0pt} {\mathcal L}_{\rm prop} 
 = - \! \sum^{N}_{n=1} \sum^{T}_{t=1} g_{n} \! &{\Big \{} z_{n,t} \log \sigma(y_{n,t}) \! 
\nonumber\\[-1pt]
&+ ~ \! (1 \! - \! z_{n,t})\log {\big (} 1 \! - \! \sigma(y_{n,t}) {\big )} {\Big \}},
\label{eq:proposed_loss_01}
\vspace{-10pt}
\end{align}

\noindent where $g_{n}$ is a gate function that controls the weight of training of two types of events.
More specifically, the gate function is calculated as

\vspace{-15pt}
\begin{align}
\vspace{-15pt}
\hspace*{0pt} g_{n}=\alpha_{s} f_{n}+ (1-\alpha_{s} ) (1-f_{n}), 
\vspace{-15pt}
\end{align}

\noindent where $\alpha_{s}$ is a progressive parameter, which is changed from 0 to 1 with time-step {\it s} (epoch) during training. 
$f_{n}$ is an event-flag.
If an event $n$ occurs at least once in the acoustic scene of the input audio, the frag is 1; otherwise, it is 0.

As shown in Fig. \ref{fig:state_learning}, in the early stage of the training, only the events that do not occur in an acoustic scene are trained.
On the other hand, in the late stage of the training, only the events that are present in an acoustic scene are trained.
Note that whether an event is difficult- or easy-to-train is determined by each acoustic scene label of an audio clip.
For example, a dataset includes a scene {\tt A}.
Events {\tt a} and {\tt b} occur at least once in the scene {\tt A}.
An event {\tt c} does not occur in the scene {\tt A}.
When the scene label of input audio is {\tt A},  {\tt a} and {\tt b} are regarded as the difficult-to-train events.
{\tt c} is regarded as the easy-to-train event.

%
%
\begin{table}[t]
\small
\captionsetup{labelformat=empty,labelsep=none}
\caption{{\bf Table 1}. Experimental conditions}
\vspace{-5pt}
\label{tbl:parameter}
\centering
\scalebox{1.00}[1.00]{
\begin{tabular}{ll}
\wcline{1-2}
&\\[-7pt]
Acoustic feature & Log-mel energy (64 dim.)\\
Frame length \hspace{-3pt} / \hspace{-3pt} shift & 40 ms \hspace{-3pt} / \hspace{-3pt} 20 ms\\
Length of sound clip & 10 s\\\hline
&\\[-8pt]
Network architecture & 3 CNN + 1BiGRU + 1 fully con.\\
\# channels of CNN layers & 128, 128, 128 \\
Filter size & 3$\times$3 \\
Pooling size & 8$\times$1, 2$\times$1, 2$\times$1 (max pooling) \\
&\\[-9pt]
\# units in GRU layer & 32 \\
\# units in fully con. layer & 32 \\
\# units in output layer & 25 \\
Threshold & 0.5 \\
\wcline{1-2}
\end{tabular}
}
\vspace{-5pt}
\end{table}
\begin{table*}[t]
\small
\captionsetup{labelformat=empty,labelsep=none}
\caption{\textbf{Table 3}. SED performance for each event}
\vspace{-18pt}  
\label{tbl:each_event}
\begin{center}
\scalebox{0.80}[0.80]{
\begin{tabular}{llcccccccccccc}
\wcline{1-14}
&&&&&&&&&&&&&\\[-8pt]
\multicolumn{2}{c}{\multirow{2}{*}{Event}} & (object) & (object) & (object) & (object) & (object) & bird & brakes & \multirow{2}{*}{breathing} & \multirow{2}{*}{car} & \multirow{2}{*}{children} & \multirow{2}{*}{cupboard} & \multirow{2}{*}{cutlery}\\[-1pt]
&& banging & impact & rustling & snapping &  squeaking & singing & squeaking &&&&&\\
&&&&&&&&&&&&&\\[-20pt]
\\\hline
&&&&&&&&&&&&&\\[-10pt]
\multirow{4}{*}{BCE} & \multirow{2}{*}{F-score} & 0.00\% & 0.37\% & \bf 6.06\% & 0.00\% & 0.00\% & 46.31\% & \bf 1.68\% & 0.00\% & 43.46\% & 0.00\% & 0.00\% & 0.00\% \\
&&&&&&&&&&&&&\\[-12pt]
& & \scriptsize$\pm$0.00 & \scriptsize$\pm$0.64 & \scriptsize$\pm$8.85 & \scriptsize$\pm$0.00 & \scriptsize$\pm$0.00 & \scriptsize$\pm$11.56 & \scriptsize$\pm$5.73 & \scriptsize$\pm$0.00 & \scriptsize$\pm$6.19 & \scriptsize$\pm$0.00 & \scriptsize$\pm$0.00 & \scriptsize$\pm$0.00  \\\cline{2-14}
&&&&&&&&&&&&&\\[-10pt]
 & \multirow{2}{*}{Error rate} & 1.00 & \bf 1.01 & \bf 0.99 & 1.00 & 1.00 & \bf 0.91 & 0.99 & 1.00 & \bf 1.06 & 1.00 & 1.00 & 1.00\\
&&&&&&&&&&&&&\\[-12pt]
& & \scriptsize$\pm$0.00 & \scriptsize$\pm$0.01 & \scriptsize$\pm$0.03 & \scriptsize$\pm$0.00 & \scriptsize$\pm$0.00 & \scriptsize$\pm$0.05 & \scriptsize$\pm$0.03 & \scriptsize$\pm$0.00 & \scriptsize$\pm$0.06 & \scriptsize$\pm$0.00 & \scriptsize$\pm$0.00 & \scriptsize$\pm$0.00 \\\hline
& \multirow{2}{*}{F-score} & 0.00\% & \bf 0.78\% & 0.77\% & 0.00\% & 0.00\% & \bf 48.88\% & 1.46\% & 0.00\% & \bf 44.83\% & 0.00\% & 0.00\% & 0.00\% \\
&&&&&&&&&&&&&\\[-12pt]
Proposed& & \scriptsize$\pm$0.00 & \scriptsize$\pm$1.15 & \scriptsize$\pm$1.70 & \scriptsize$\pm$0.00 & \scriptsize$\pm$0.00 & \scriptsize$\pm$5.55 & \scriptsize$\pm$3.86 & \scriptsize$\pm$0.00 & \scriptsize$\pm$7.50 & \scriptsize$\pm$0.00 & \scriptsize$\pm$0.00 & \scriptsize$\pm$0.00  \\\cline{2-14}
method & \multirow{2}{*}{Error rate} & 1.00 & 1.02 & 1.03 & 1.00 & 1.00 & 0.96 & 0.99 & 1.00 & 1.15 & 1.00 & 1.00 & 1.00\\
&&&&&&&&&&&&&\\[-12pt]
& & \scriptsize$\pm$0.00 & \scriptsize$\pm$0.03 & \scriptsize$\pm$0.05 & \scriptsize$\pm$0.00 & \scriptsize$\pm$0.00 & \scriptsize$\pm$0.09 & \scriptsize$\pm$0.02 & \scriptsize$\pm$0.00 & \scriptsize$\pm$0.23 & \scriptsize$\pm$0.00 & \scriptsize$\pm$0.00 & \scriptsize$\pm$0.00 \\\hline
&&&&&&&&&&&&&\\[-12pt]
\wcline{1-14}
\end{tabular}
}
\end{center}
\vspace{-14pt}
\small
\label{tbl:each_event}
\begin{center}
\scalebox{0.78}[0.78]{
\begin{tabular}{llccccccccccccc}
\wcline{1-15}
&&&&&&&&&&&&&\\[-8pt]
\multicolumn{2}{c}{\multirow{2}{*}{Event}} & \multirow{2}{*}{dishes} & \multirow{2}{*}{drawer} & \multirow{2}{*}{fan} & glass & keyboard & large & mouse & mouse & people & people & washing & water tap & wind
\\[-1pt]
&&  &  &  & jingling & typing & vehicle & clicking & wheeling & talking & walking & dishes & running & blowing\\
&&&&&&&&&&&&&\\[-20pt]
\\\hline
&&&&&&&&&&&&&&\\[-10pt]
\multirow{4}{*}{BCE} & \multirow{2}{*}{F-score} & 0.00\% & 0.00\% & 9.95\% & 0.00\% & 0.00\% & 16.93\% & 0.00\% & 0.00\% & 0.00\% & 2.67\% & 17.25\% & 40.78\% & \bf 0.55\% \\
&&&&&&&&&&&&&&\\[-12pt]
& & \scriptsize$\pm$0.00 & \scriptsize$\pm$0.00 & \scriptsize$\pm$18.74 & \scriptsize$\pm$0.00 & \scriptsize$\pm$0.00 & \scriptsize$\pm$1.43 & \scriptsize$\pm$0.00 & \scriptsize$\pm$0.00 & \scriptsize$\pm$0.08 & \scriptsize$\pm$3.66 & \scriptsize$\pm$12.70 & \scriptsize$\pm$12.36 & \scriptsize$\pm$0.21 \\\cline{2-15}
& \multirow{2}{*}{Error rate} & 1.00 & 1.00 & 0.95 & 1.00 & 1.00 & 6.17 & 1.00 & 1.00 & 1.06 & 1.02 & \bf 1.19 & \bf 0.83 & \bf 1.00 \\
&&&&&&&&&&&&&&\\[-12pt]
& & \scriptsize$\pm$0.00 & \scriptsize$\pm$0.00 & \scriptsize$\pm$0.13 & \scriptsize$\pm$0.00 & \scriptsize$\pm$0.00 & \scriptsize$\pm$0.67 & \scriptsize$\pm$0.00 & \scriptsize$\pm$0.00 & \scriptsize$\pm$0.09 & \scriptsize$\pm$0.03 & \scriptsize$\pm$0.17 & \scriptsize$\pm$0.10 & \scriptsize$\pm$0.00 \\\hline
& \multirow{2}{*}{F-score} & \bf 1.22\% & 0.00\% & \bf 50.63\% & 0.00\% & \bf 0.01\% & \bf 18.13\% & 0.00\% & 0.00\% & \bf 0.06\% & \bf 3.43\% & \bf 24.80\% & \bf 45.07\% & 0.07\% \\
&&&&&&&&&&&&&&\\[-12pt]
Proposed & & \scriptsize$\pm$3.15 \scriptsize$\pm$0.00 & \scriptsize$\pm$26.68 & \scriptsize$\pm$0.00 & \scriptsize$\pm$0.03 & \scriptsize$\pm$5.02 & \scriptsize$\pm$0.00 & \scriptsize$\pm$0.00 & \scriptsize$\pm$0.17 & \scriptsize$\pm$3.46 & \scriptsize$\pm$13.15 & \scriptsize$\pm$4.82 & \scriptsize$\pm$0.20 \\\cline{2-15}
method & \multirow{2}{*}{Error rate} & 1.00 & 1.00 & \bf 0.77 & 1.00 & 1.00 & \bf 4.54 & 1.00 & 1.00 & \bf 1.03 & 1.02 & 1.32 & 0.84 & 1.01 \\
&&&&&&&&&&&&&&\\[-12pt]
& & \scriptsize$\pm$0.01 & \scriptsize$\pm$0.00 & \scriptsize$\pm$0.19 & \scriptsize$\pm$0.00 & \scriptsize$\pm$0.00 & \scriptsize$\pm$2.27 & \scriptsize$\pm$0.00 & \scriptsize$\pm$0.00 & \scriptsize$\pm$0.03 & \scriptsize$\pm$0.03 & $\pm$\scriptsize0.26 & $\pm$\scriptsize0.04 & \scriptsize$\pm$0.02 \\
&&&&&&&&&&&&&&\\[-12pt]
\wcline{1-15}
\end{tabular}
}
\vspace{-10pt}
\end{center}
\end{table*}
\begin{table}[t]
\small
\captionsetup{labelformat=empty,labelsep=none}
\caption{\textbf{Table 2}. Overall performance of SED}
\vspace{-18pt}
\label{tbl:F-score}
\begin{center}
\scalebox{0.88}[0.88]{
\begin{tabular}{ccccccc}
\wcline{1-7}
&&&&&&\\[-7pt]
\multicolumn{3}{c}{\multirow{2}{*}{Method}} & \multicolumn{2}{c}{F-score} & \multicolumn{2}{c}{Error rate} \\
&&&&&&\\[-13pt]
&&& micro & macro & micro & macro \\\hline
&&&&&&\\[-9pt]
\multicolumn{3}{l}{\multirow{2}{*}{BCE}} & 25.30\% & 7.44\% & 1.00 & 1.21 \\[-2pt]
&&&&&&\\[-13pt]
&&& \scriptsize$\pm$4.72 & \scriptsize$\pm$1.21 & \scriptsize$\pm$0.04 & \scriptsize$\pm$0.03\\[0pt]
\multicolumn{3}{l}{\multirow{2}{*}{MTL of SED \& SAD}} & 26.62\% & 7.36\% & 1.02 & 1.20 \\[-2pt]
&&&&&&\\[-13pt]
&&& \scriptsize$\pm$2.68 & \scriptsize$\pm$0.65 & \scriptsize$\pm$0.03 & \scriptsize$\pm$0.09\\[0pt]
\multicolumn{3}{l}{\multirow{2}{*}{MTL of SED \& ASC}} & 26.12\% & 7.46\% & 0.97 & 1.18\\[-2pt]
&&&&&&\\[-13pt]
&&& \scriptsize$\pm$3.94 & \scriptsize$\pm$0.58 & \scriptsize$\pm$0.07 & \scriptsize$\pm$0.06 \\[0pt]\hdashline
&&&&&&\\[-8pt]
\multicolumn{3}{l}{\multirow{2}{*}{Proposed method}} & \bf 35.39\% & \bf 9.61\% & \bf 0.85 & \bf 1.15\\[-4pt]
&&& \scriptsize$\pm$6.06 & \scriptsize$\pm$1.24 & \scriptsize$\pm$0.07 & \scriptsize$\pm$0.09 \\[0pt]
\multicolumn{3}{l}{\multirow{2}{*}{Proposed+MTL of SED \& SAD}} &  \bf 33.57\% & \bf 9.11\% & \bf 0.93 & \bf 1.17\\[-4pt]
&&& \scriptsize$\pm$4.86 & \scriptsize$\pm$0.81 & \scriptsize$\pm$0.07 & \scriptsize$\pm$0.07 \\[0pt]
\multicolumn{3}{l}{\multirow{2}{*}{Proposed+MTL of SED \& ASC}} &  \bf 35.62\% & \bf 9.65\% & \bf 0.85 & \bf 1.15\\[-4pt]
&&& \scriptsize$\pm$6.35 & \scriptsize$\pm$1.31 & \scriptsize$\pm$0.09 & \scriptsize$\pm$0.09 \\[0pt]\wcline{1-7}
\end{tabular}
}
\vspace{-22pt}
\end{center}
\end{table}
%

\vspace{-5pt}
\section{Experiments}
\label{sec:exp}
\vspace{-2pt}
\subsection{Experimental conditions}
\label{sec:condition}
\vspace{-2pt}
To evaluate the performance of the proposed method, we conducted evaluation experiments using the TUT Sound Events 2016 \cite{Mesaros2016TUTDF}, TUT Sound Events 2017 \cite{Mesaros2017}, TUT Acoustic Scenes 2016 \cite{Mesaros2016TUTDF}, and TUT Acoustic Scenes 2017 \cite{Mesaros2017} datasets.
From these datasets, we selected sound clips including four acoustic scenes, ``home,'' ``residential area'' (TUT Sound Events 2016), ``city center'' (TUT Sound Events 2017, TUT Acoustic Scenes 2017), and ``office'' (TUT Acoustic Scenes 2016), which contain 266 min (development set, 192 min; evaluation set, 74 min) of audio.
Here, the acoustic scene ``office'' in TUT Acoustic Scenes 2016 and ``city center'' in TUT Acoustic Scenes 2017 did not have sound event labels.
We thus manually annotated the sound clips with sound event labels by the procedure described in \cite{Mesaros2016TUTDF,Mesaros2017}.
These sound clips include the 25 types of sound event labels.
Fig. \ref{fig:frames_events} shows the numbers of active time frames of sound events on the development set that we used.
The labels of events annotated for our experiment are available in \cite{Imoto_dataset2019_01}.

As acoustic features, we used 64-dimensional log-mel energies calculated for each 40 ms time frame with 50\% overlap.
This setting is from the baseline system of the DCASE2018 Challenge task4 \cite{setting01}.
As the baseline model of SED, we used the convolutional neural network and bidirectional gated recurrent unit (CNN--BiGRU) \cite{SED_CRNN}.
Moreover, to verify the usefulness of the proposed method, we used a model combining SED and sound activity detection (SAD) based on multitask learning (MTL), referred to as ``MTL of SED \& SAD'' \cite{SED_SAD}, and a model combining SED and ASC, referred to as ``MTL of SED \& ASC'' \cite{Tonami_WASPAA2019_01}.
The sound activity detection is the mechanism of recognizing any active events in a time frame.
The reason why we choose MTL of SED \& SAD is that this modern method, in which no information on the scene is considered, is simple but effective.
MTL of SED \& ASC is the multitask-learning-based SED with ASC, which uses scene labels by ASC.
Other experimental conditions are listed in Table~\ref{tbl:parameter}.
In Table~\ref{tbl:parameter}, X $\times$ Y denotes that the filter size is X along the frequency axis by Y along the time axis.
As the evaluation metric, the frame-based metric is used.
We conduct the experiments using ten initial values.
To evaluate the SED performance, a segment-based metric \cite{Mesaros2016_MDPI} is used.
In this work, the size of a segment is set to the frame length.

In this work, we adopt the following exponential scheduler as the progressive parameter in Eq. 3:

\vspace{-10pt}
\begin{align}
\vspace{-15pt}
 \hspace*{0pt} \alpha_{s} = \biggl( \frac{s}{s_{max}} \biggl)^\lambda , 
\vspace{-15pt}
\end{align}
\vspace{-10pt}

\noindent where $s$ and $s_{max}$ represent the current and maximum epoch, respectively.
$\lambda$ is tuned using the development dataset and is set as 2.0.  
%
%
\vspace{-8pt}
\subsection{Experimental results}
\label{sec:results}
\vspace{-3pt}
Table 2 shows the SED performances in terms of the segment-based F-score and error rate.
In Table 2, micro and macro indicate the overall and class-average scores, respectively.
The numbers to the right of $\pm$ represent standard deviations.
``BCE'' is the CNN--BiGRU using the BCE loss. 
``Proposed method'' represents the SED performance using Eqs. 2 and 3 with CNN--BiGRU.
``Proposed+MTL of SED \& SAD'' indicates the SED performance using Eqs. 2 and 3 with SAD.
``Proposed+MTL of SED \& ASC'' denotes the multitask-learning-based SED with ASC using the proposed objective function for SED.
The results show that the proposed method achieves a more reasonable performance than the conventional BCE.
Moreover, when using SAD and the model combining SED and ASC with the proposed objective function, the SED performance is better than those of the conventional MTL of SED \& SAD and the MTL of SED \& ASC.
In particular, ``Proposed method'' improves the F-score of SED by 10.09 percentage points compared with that of the conventional SED using the BCE.
The results indicate that the proposed method considering the  training difficulty of events enables  more effective SED performances than the conventional method using the BCE.

To investigate in detail the SED performance, we observed the segment-based F-score and error rate for each event.
Table 3 indicates the SED performance for each event.
As shown in Table 3, the proposed method outperformed the conventional SED using the BCE for many events.
In particular, the F-scores for ``fan,'' ``washing dishes,'' and ``water tap running'' are more significantly improved in the proposed method than in the conventional method.
This might be because the active frames of these events occur continuously, that is, these events are relatively effortless to be detected compared with other events. 
On the other hand, the F-scores for ``(object) rustling,'' ``brakes squeaking,'' and ``wind blowing'' do not improve.
This is because the numbers of the events of the active frames are too small as shown in Fig. \ref{fig:frames_events}.
In other words, the active frames are trained mainly in the late stage of training when using the proposed method.
This may also lead to the poor results for some events when using the proposed method.

\vspace{-12pt}
\section{Conclusion}
\label{sec:conc}
\vspace{-8pt}
In this paper, we proposed the curriculum-learning-based objective function for SED.
In the proposed method, we applied the training difficulty between sound events considering acoustic scenes to the conventional BCE loss.
More specifically, the SED models using the proposed method are trained from the easy-to-train to difficult-to-train events during training. 
The experimental results indicate that the proposed method improves the F-score of the SED by 10.09 percentage points compared with that of the conventional CNN--BiGRU using the BCE loss.
In our future work, we will investigate a more effective method for SED considering the relationship between sound events and acoustic scenes.

\vspace{-4pt}
\section{Acknowledgement}
\label{sec:ack}
\vspace{-4pt}
This work was supported by JSPS KAKENHI Grant Number JP19K20304.
\vspace{-8pt}
%
\begin{spacing}{0.99}
\bibliographystyle{IEEEbib}
\bibliography{refs}
\end{spacing}

\end{document}